\documentclass[aps,prb,amsmath,amssymb,twocolumn]{revtex4}

\usepackage{graphicx}

\begin{document}
\title{Dissipation, topology,  and quantum phase transition  in a one-dimensional Joesphson junction array}
\author{Pallab Goswami}
\affiliation{Department of Physics and Astronomy\\ University of California Los Angeles, Los Angeles, California, 90095-1547} 
\author{Sudip Chakravarty}
\affiliation{Department of Physics and Astronomy\\ University of California Los Angeles, Los Angeles, California, 90095-1547} 
\date{\today}

\begin{abstract}
 We study the phase diagram and quantum critical properties of a resistively shunted Josephson junction array in one dimension from a strong coupling analysis. After mapping the dissipative quantum phase model to an effective sine-Gordon model we study 
the renormalization group flow and  the phase diagram. We try to bridge the phase 
diagrams obtained from the weak and the strong coupling renormalization group calculations to extract a more comprehensive picture of the complete 
phase diagram. The relevance of our theory to experiments in nanowires is discussed.
\end{abstract}
\pacs{PACS numbers:}

\maketitle

\section{Introduction }

Ohmic dissipation was shown to be of critical importance in the macroscopic quantum coherence of a double-well system.\cite{Chakravarty,Bray} This zero-dimensional problem undergoes a spontaneous symmetry-breaking transition in the ground state if the resistor causing the dissipation is smaller than the  quantum of resistance $R_{Q}=h/4e^{2}\approx 6.5\textrm{k}\Omega$; $h$ is the Planck's constant and $e$ the electronic charge. It is perhaps the simplest example of a quantum critical point (QCP).\cite{Subir} Soon afterwards, it was also shown\cite{Schmid} that a  resistively shunted Josephson junction undergoes a similar phase transition from a metallic state (because there is always an Ohmic shunt present in the model), in which the phase difference between the superconductors  is delocalized,  to a superconducting state (in which a supercurrent can flow below the critical current), where the phase difference is localized in a minimum of the cosine potential of the effective classical Hamiltonian derived by Josephson. The root of this remarkable QCP is logarithmic interactions between the instantons\cite{Coleman} representing the tunneling events, resulting in a dramatic failure of the dilute instanton gas approximation, so successful in the macroscopic quantum tunneling problem.\cite{Caldeira,Chang} The only difference between the two systems is the specific ordering of the instantons,\cite{Coleman} but in both cases the underlying cause is the orthogonality catastrophe\cite{Anderson} of an infrared divergent heat-bath that is necessary to model the Ohmic shunt resistor. Since then, numerous suggestions have been made that such an essentially dynamic zero temperature  ($T=0$) phase transition may be embedded in many condensed matter systems, and this may explain the ubiquity of dissipation in many ultra low temperature phenomena. Instead of enumerating here an incomplete list of references, we refer to Ref.~\onlinecite{Kapitulnik}.

What about experiments regarding these predictions? We do not merely mean a reduction of the quantum tunneling rate due to dissipation, but a sharp phase transition or a QCP. Surprisingly, experimental evidence is sparse. The closest experiment that indicates symmetry breaking in a double well is an indirect experiment involving a superconducting quantum interference device.\cite{Lukens2} In contrast, in a more direct experiment on a single resistively shunted Josephson junction  such a dynamic phase transition has been apparently experimentally observed. \cite{Penttila}. However, it seems that the theoretical situation is more subtle than that assumed in the past.\cite{Werner} An attempt to observe such a transition in superconducting nanowires was also made.\cite{Bezryadin1} Unfortunately, the present experimental situation appears to be unclear.\cite{Bezryadin2,Bezryadin3,Bezryadin4,Chan}

Sometimes the simplest theoretical concept is not the simplest from the experimental perspective. From the very beginning it was realized that such a dynamic transition may have important consequences in many body problems \cite{Chakravarty1,Chakravarty2,MPAF1,MPAF2}, in particular in resistively shunted Josephson junction arrays (RSJJA). Beginning with the pioneering work of Orr {\em et al.}\cite{Orr} experiments have been few and far in between.
A brief recent review  is provided by Goldman.\cite{Goldman} Of importance to us here are the experiments of Rimberg {\em et al.},\cite{Rimberg} Takahide {\em et al.},\cite{Takahide} and Miyazaki {\em et al.}\cite{Miyazaki} On the theoretical side, many important contributions have been made, but we list here only the papers that are germane to our present work; these are Refs.~\onlinecite{Panyukov, Bobbert1, Bobbert2, Korshunov1, Korshunov2, Zwerger1, Zwerger2} .

RSJJA is a simple but nontrivial model, almost a paradigm to use a well-worn word,  where both dynamics and statics can simultaneously play an important role in a quantum phase transition, as opposed to a classical phase transition.  Compared to a dissipative single junction problem, RSJJA  is theoretically more challenging because of the interplay of both spatial and temporal fluctuations. Indeed  it is shocking that a recent theoretical work\cite{Tewari1,Tewari2} has found that 
the ground state of RSJJA is a state where the state of  the $(0+1)$-dimensional
elements (single Josephson junctions) can slide past each other
despite couplings between them. This is despite arbitrarily long-ranged spatial couplings. Such a phase, called the floating phase, was derived from a renormalization group analysis that is perturbative in the Josephson coupling. Given the striking nature of a lower-dimensional quantum criticality embedded in a higher dimensional manifold, it behooves us to examine the phenomena in the strong coupling limit and to see how the two limits are reconciled. From the earlier hints,\cite{Chakravarty2} the strong coupling analysis should show that when the Josephson coupling is gradually increased, the quantum phase transition changes its nature and cease to be entirely dynamic---topology, quantum mechanics, dissipation, and the collective nature of the problem, all become equally important unlike the weak coupling limit.
Motivated by the theoretical challenge,  recent experiments, and the paradigmatic nature of RSJJA, we analyze it in one dimension from the strong coupling limit. As the conceptually important issues are already present in the one-dimensional case, there is no need to examine more complicated higher dimensional situations. 

We use Villain mapping to investigate the strong coupling phase diagram and the critical properties. Villain mapping has been used previously  to map the phase slip processes to a neutral gas of charges that  have anisotropic logarithmic interactions in imaginary time and spatial directions because of dissipation. \cite{Bobbert1, Bobbert2, Zwerger1, Zwerger2}We improve upon the previous work and show that the resulting phase diagram is remarkably different. In a Josephson junction array in the absence of dissipation, quantum phase slips are the topological excitations whose classical counterparts are vortices of the classical XY model. One of the basic concepts behind this mapping is a conservation law. Due to tunneling, the number of Cooper pairs will change on the superconducting grains. In the absence of any source or a sink this leads to a continuity equation on the lattice which serves as a constraint on the integer tunneling current. This constraint is resolved in terms of a single integer field defined  on a dual space-time lattice, resulting in familiar vortices. Thus, the superconductor-insulator transition due to unbinding of vortices can be described by a one component sine-Gordon model. The presence  of  shunt resistances, or dissipation, alters this picture. Since we need to include the current through the shunts in addition to the tunneling current,  we get a different conservation law with a source term (This point was missed in the past.) The presence of the source term does not allow us to resolve the current constraint in terms of a single integer field on the dual lattice, and we are left with a neutral gas  of two-flavored charges and anisotropic long range interactions in space-time. Consequently, our phase diagram is different from those derived in the past. 

The renormalization group equations, derived here for the first time, are obtained by mapping the two-component neutral charge gas system to a two-component sine-Gordon model. A two component sine-Gordon model was used previously by Refael {\em et al} in the context of  two resistively shunted Josephson junctions.\cite{Refael}   There are therefore some  similarities in the phase diagrams. Apart from fully superconducting and metallic states some partially ordered phases are found. Surprisingly, the two-component sine-Gordon model also arises in the context of a classical two-dimensional XY model in the presence of an in-plane magnetic field. \cite{Fertig} The  magnetic field spoils the zero divergence condition by introducing a source term. Though it is clear that RSJJA and XY model in an in-plane magnetic field describe different physics, there are some similarities between the RG analyses of both problems.

Our paper is organized as follows. In Sec. II we describe the microscopic model and provide briefly the weak coupling results. In Sec III we perform a detailed strong coupling analysis employing Villain mapping. To perform RG calculations we map the two flavored neutral charge gas problem arising from Villain mapping to a two component sine-Gordon model. In Sec. IV we describe the fixed point analysis
of the RG recursion relations. In Sec. V we construct the strong coupling phase diagram and contrast the results with those of the previous authors. In Sec VI we briefly mention the relation between  a superconducting nanowire and a RSJJA . Finally we summarize our results in Sec VII. We present all the technical details of the RG calculations for the two component sine-Gordon model in the Appendix. Althogh this is relegated to the Appendix to preserve a smooth flow of the text, this Appendix is the heart of our theoretical work.

\section{The Model}
We consider the quantum action given by
\begin{eqnarray}
S &=& \int_{0}^{\beta}d\tau \sum_{i}\frac{\dot{\theta}_{i}^{2}}{2E_{0}}+\frac{\beta\alpha}{4\pi}\sum_{{i,n}}\mid\omega_{n}\mid\mid\partial_{x}\theta_{i}(n)\mid^{2} \nonumber \\
    &  &-\int_{0}^{\beta}d\tau V\sum_{{i}}\cos(\partial_{x}\theta_{i})
\end{eqnarray}
$\beta$ is the inverse temparature, $E_{0}=2e^{2}/C$ is the charging energy, where C is the capaciatance of single island. $\omega_{n}$ is the Matsubara frequency and $\alpha=h/4e^{2}R$ is the dissipation strength, where R is the shunt resistance. $\partial_{x}\theta_{i}=\theta_{i}-\theta_{i+1}$, and V is the Josephson coupling  between the grains.

In the weak coupling regime ($\frac{V}{E_{0}}\ll1$), which is dominated by strong quantum fluctuations, RG calculations lead to the following recursion relations upto order $O(V^{2})$ 
 \begin{eqnarray}
 \frac{dV}{dl} & = & (1-\frac{1}{\alpha})V \nonumber \\
 \frac{d\alpha}{dl} & = & 0
 \end{eqnarray}
So, $\alpha =1$ is a fixed line in the weak coupling limit. For $\alpha>1$, system is  ordered (superconducting), and $V$ is a relevant operator. For $\alpha<1$ the system is disordered (metallic),  and $V$ is irrelevant. Recently Tewari {\em et al.} \cite{Tewari1, Tewari2} has extended this calculation to the order $O(V^{3})$ and have shown that for a one-dimensional array the above set 
 of RG equations are correct upto third order. They also demonstrated that including longer ranged  couplings that when $V$ is irrelevant all longer range  Josephson couplings are also irrelevant.  They become simultaneously relevant with $V$ when $\alpha>1$. So, around the critical line $\alpha =1$, the chain of Josephson  junctions decouples and behaves as independent single junctions. For these reasons the disordered phase (metallic) has been characterized as a floating phase. It is important to understand how far in the $V-\alpha$ plane is this weak coupling RG picture valid.

\section{A Strong coupling analysis}
 In the  strong coupling limit, that is $V/E_{0}\gtrsim 1$, the  phase difference between the neighboring grains will be localized around the minima of the periodic cosine potential and the tunneling events between the minima will govern the low energy physics. We analyze these tunneling events using Villain mapping.
To employ Villain mapping we first discretize the imaginary time into $N$ slices so that $\beta=N\Delta\tau$ where $\Delta\tau$ is the lattice spacing in imaginary time direction. After discretizing, the partition function becomes
\begin{eqnarray}
Z  & = & \int[D\theta]\exp[J_{\mu}\sum_{{\vec{l}}}(1-\cos\nabla_{\mu}\theta(\vec{l})) \nonumber \\
    &     &-\sum_{{\vec{q}}}\frac{\alpha}{4\pi}\mid\omega \mid \Delta \tau f(k)\mid \theta(\vec{q})\mid^{2}],
\end{eqnarray}
where the summation over repeated indices is implied. $\mu\equiv (\tau, x)$ represents the components on the space-time lattice and $\vec{l}$ stands for the space-time coordinates of the lattice points. Here $J_{\tau}=1/E_{0}\Delta \tau$, $J_{x}=V\Delta \tau$ and $f(k)=2[1-\cos(ka)]$. For both $J_{\tau}$ and $J_{x}$ large we can replace the cosine by a periodic Gaussian, as follows:
\begin{eqnarray}
Z & = & \int[D\theta]\sum_{m_{\mu}(\vec{l})}\exp[-\frac{J_{\mu}}{2}\sum_{{\vec{l}}}(\nabla_{\mu}\theta(\vec{l})-2\pi m_{\mu}(\vec{l}))^{2} \nonumber \\
   &     &-\sum_{{\vec{q}}}\frac{\alpha}{4\pi}\mid\omega \mid \Delta \tau f(k)\mid \theta(\vec{q})\mid^{2}].
   \label{eq:Z}
\end{eqnarray}
Here $m_{\mu}(i,\tau)$ are integers and represent tunneling between minima of the cosine potential. 

Before we proceed further with the action described by Eq.~\ref{eq:Z}, we explicitly demonstrate the difference of the conservation laws in non-dissipative and a dissipative Josephson junction array as have been mentioned in the Introduction. First we choose the non-dissipative case, and we set $\alpha=0$ in Eq.  4. We also introduce two auxiliary fields $B_{\mu}$ to decouple the periodic Gaussian terms as
\begin{eqnarray}
Z \propto  \int[D\theta] [D B_{\mu}]\sum_{m_{\mu}(\vec{l})}\exp[-\sum_{\mu, \vec{l}}(\frac{B^{2}_{\mu}(\vec{l})}{2J_{\mu}}+iB_{\mu}(\vec{l}) \times \nonumber \\
(\nabla_{\mu}\theta(\vec{l})-2\pi m_{\mu}(\vec{l})))]
\end{eqnarray}
Summation over the integer field $m_{\mu}$ restricts the continuous fields $B_{\mu}$ to integer values
and integrating out $\theta$ we obtain
\begin{equation}
Z \propto \sum_{B_{\mu}} \delta_{\nabla_{\mu}B_{\mu},0}\exp[-\sum_{\mu, \vec{l}}\frac{B^{2}_{\mu}(\vec{l})}{2J_{\mu}}]
\end{equation}
So, we obtained the conservation law $\nabla_{\mu}B_{\mu}=0$ in the form of a constraint on the configurations of $B_{\mu}$. If we choose $B_{x}=\partial_{\tau}h, B_{\tau}=-\partial_{x}h$ where $h$ is a single component integer field the constraint is resolved. For any nonzero $\alpha$, we introduce an additional auxiliary field $\rho$ to decouple the quadratic dissipative term:
\begin{eqnarray}
& &\exp[-\sum_{{\vec{q}}}\frac{\alpha}{4\pi}\mid\omega \mid \Delta \tau f(k)\mid \theta(\vec{q})\mid^{2}] 
\nonumber \\
& &=\int[D\rho]\exp[-\sum_{\vec{q}}\frac{|\rho(\vec{q})|^{2}}{(\frac{\alpha}{4\pi}\mid\omega \mid \Delta \tau f(k))}+i\sum_{\vec{l}}\rho(\vec{l})\theta(\vec{l})]. \nonumber \\
\end{eqnarray}
Now integrating out $\theta$
we obtain the modified conservation law
\begin{equation}
\nabla_{\mu}B_{\mu}+\rho=0
\end{equation}
which directly demonstrates that the zero divergence condition is destroyed and the auxiliary field $\rho$ serves as a source term. 

We proceed with an analysis of Eq.~\ref{eq:Z}. Integrating out $\theta$ from the partition function and doing the following transformations,
\begin{eqnarray}
p_{1}(i,\tau) & = & m_{\tau}(i+1,\tau)-m_{\tau}(i,\tau),  \\
p_{2}(i,\tau) & = & m_{x}(i,\tau +\Delta \tau)-m_{x}(i,\tau),
\end{eqnarray}
we obtain 
\begin{equation}
Z  = Z_{SW}\sum_{p^{\mu}(\vec{l})}\exp[-2\pi^{2}\sum_{{\vec{q}}}p^{\mu}(\vec{q})G_{\mu,\nu}(\vec{q})p^{\nu}(-\vec{q})].
\end{equation}
In the above step we have rescaled time direction as $\Delta \tau^{'}=\Delta \tau \sqrt{VE_{0}}a$.  So, we have rewritten the partition function in terms of two kinds of interacting charges $p_{1}$ and $p_{2}$ which by construction form a neutral gas of charges. Interactions of the charges are encoded in the $G_{\mu,\nu}(\vec{q})$ which are given by
\begin{eqnarray}
G_{11}(k,\omega^{'}) & = & \frac{J}{\Delta \tau^{'}a (k^{2}+\omega^{'2})}+\frac{\alpha}{2\pi\Delta \tau^{'}}\frac{\mid \omega^{'}\mid}{(k^{2}+\omega^{'2})}, \\
G_{22}(k,\omega^{'}) & = & \frac{J}{\Delta \tau^{'}a (k^{2}+\omega^{'2})}+\frac{\alpha}{2\pi}\frac{1}{\omega^{'} \Delta \tau^{'}}\nonumber \\
                                      &     &-\frac{\alpha}{2\pi}\frac{1}{\Delta \tau^{'}}\frac{\mid \omega^{'}\mid}{(k^{2}+\omega^{'2})},\\
G_{12}(k,\omega^{'}) & = & -\frac{J}{\Delta \tau^{'}a (k^{2}+\omega^{'2})}=G_{21}
\end{eqnarray}
In the long-wavelength and low-frequency limit,
\begin{equation}
G_{ij}=(-1)^{i+j}\frac{J}{\Delta \tau^{'}a (k^{2}+\omega^{'2})}+\delta_{i,2}\delta_{j,2}\frac{\alpha}{2\pi}\frac{1}{\omega^{'} \Delta \tau^{'}},
\end{equation}
where $J=\sqrt{J_{x}J_{\tau}}=\sqrt{V/E_{0}}$.
From this interaction matrix it becomes clear that because of dissipation spatial kinks,  i.e $p_{2}$,  have an additional anisotropic on-site logarithmic interaction in imaginary time in addition to isotropic logarithmic interaction in space-time. Charges of opposite sign for a given flavor attract and for different flavors charges of same sign attract. In the presence of dissipation we can not reduce the partition function in terms of a single set of charges or vortices given by $p_{3}=p_{2}-p_{1}=\epsilon_{\mu\nu}\partial_{\mu}m_{\nu}$. So, we have to work with a two-component neutral charge gas, and the corresponding sine-Gordon theory will be a two component field theory.  For convenience we shall choose $\Delta \tau=1/ \sqrt{VE_{0}}$, which leads to $\Delta \tau^{'}=a$. With this choice, the space-time lattice becomes isotropic and the anisotropy of the problem is captured only through the anisotropy of the interaction terms. 

\begin{figure}[htbp] 
   \centering
   \includegraphics[scale=0.5]{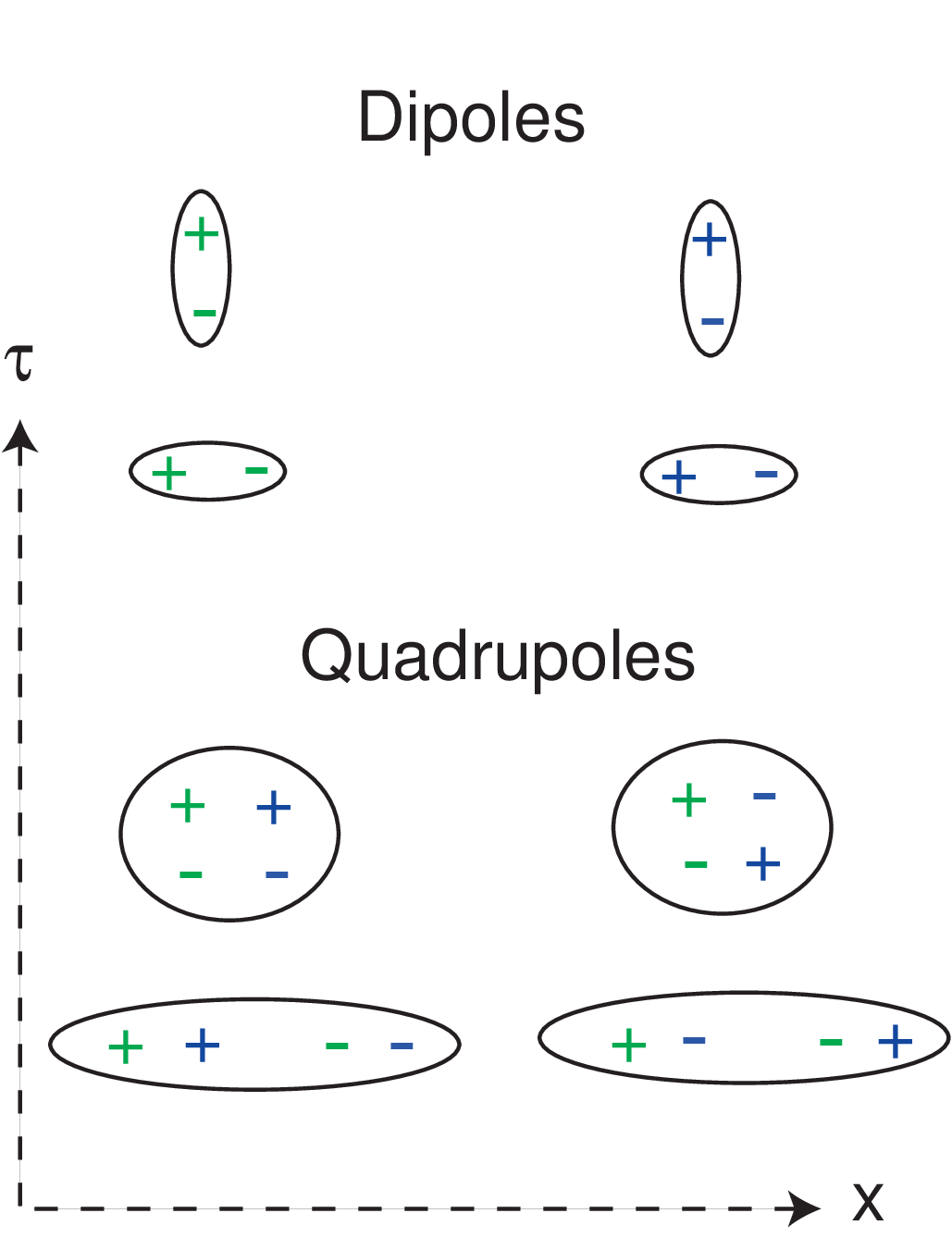} 
   \caption{Configurations of space-time dipoles and qudrupoles of different flavors. Note that equivalent pictures can be drawn by charge conjugations.}
   \label{fig:charges}
\end{figure}

We introduce two fugacities $y_{1}, y_{2}$ corresponding to $p_{1}, p_{2}$  to control the charge fluctuations.
As usual,  for small $y_{1}, y_{2}$ we can restrict $p_{1},p_{2}$ to $0,\pm1$. Thus, the sine-Gordon action is
\begin{eqnarray}
S & = &\frac{1}{2}\sum_{k,\omega}\vec{\phi}^{\ast}\widetilde{G}^{-1}\vec{\phi} -\sum_{i=1}^{2}\frac{2y_{i}}{a^{2}}\int d\vec{x}\cos(2\pi \phi_{i})\nonumber \\
  &   & -\sum_{\pm}\frac{2y_{\pm}}{a^{2}}\int d\vec{x}\cos(2\pi \phi_{1} \pm 2\pi  \phi_{2}) \end{eqnarray}
 We are using the vector notation $\vec{x}=(x,\tau)$ and $\int d\vec{x}=\int dx d\tau$. $y_{+},y_{-}$ are combinations of $y_{1},y_{2}$. At second order in fugacities, the RG procedures generate $\cos(2\pi \phi_{1} \pm 2\pi \phi_{2})$ terms, and that is the reason for extending the
coupling constant space.  $y_{i}$ controls the formation dipoles between different
signs of charges for i-th flavor. When the charges are bound as dipoles, the fugacities become irrelevant and flow to zero, implying order; when they become relevant free charges proliferate. $y_{+}$ controls
the density of dipoles between charges of same sign but of different flavors, and $y_{-}$ controls the density of dipoles between charges of opposite sign and different flavors; see, Fig.~\ref{fig:charges}. $\widetilde{G}^{-1}$ is the matrix inverse of the interaction matrix for the charges. This $2\times 2$ matrix is given by
\begin{equation}
G^{-1}_{ij}  =  \delta_{i,1}\delta_{j,1}\frac{k^{2}+\omega^{2}}{J}+\frac{2\pi}{\alpha}\mid \omega \mid 
\end{equation}

Interestingly, the presence of divergent $\frac{1}{\mid \omega \mid}$
term in the propagator requires that we formulate the RG (thinning out the degrees of freedom)  by  integrating only in a given frequency shell. Then a dimensional analysis with the scaling prescription
$k\to ke^{-l/z} , \omega \to \omega e^{-l}$, where $l$ is the length rescaling factor,  shows that 
\begin{equation}
\frac{d\alpha}{dl}=0,
\end{equation}
as $\alpha$ is dimensionless. Since, $|\omega|$ is nonanalytic and we have to maintain the periodicity of the cosine terms, the above RG equation for $\alpha$ is exact.\cite{Sachdev,Castellani}
The following recursion relations up to  second order are derived in Appendix ~\ref{app:A}
\begin{eqnarray}
\frac{dy_{1}}{dl}&=&(1+\frac{1}{z}-\pi J)y_{1}+\alpha y_{2}y_{+} +(2 \pi J + \alpha)y_{2}y_{-},  \\
\frac{dy_{2}}{dl}&=&(1+\frac{1}{z}- \pi J- \alpha)y_{2}+2 \pi Jy_{1}y_{-}, \\
\frac{dy_{+}}{dl}&=&(1+\frac{1}{z}-\alpha)y_{+}+\pi Jy_{1}y_{2}, \\
\frac{dy_{-}}{dl}&=&(1+\frac{1}{z}-2\pi J-\alpha)y_{-} - \pi J y_{1}y_{2}, \\
\frac{dJ}{dl}      &= &J[(1-\frac{1}{z})-J(A_{1}y_{1}^{2}+A_{+}y_{+}^{2}+A_{-}y_{-}^{2})].
\end{eqnarray}
To keep the coefficient of $k^{2}$ fixed we need 
\begin{equation}
\frac{1}{z}=1-\frac{A_{+}y_{+}^{2}}{2}
\end{equation}
Here $A_{1}, A_{+}, A_{-}$ are regularization dependent constants.
 
 \section{The Fixed points}
From first order RG equations, the  following picture emerges: $\frac{1}{z}=1$ and the entire
$J-\alpha$ plane is broken into six regions.  In the region $\pi J>2, \alpha>2$ all the fugacities are irrelevant, and the system is fully superconducting and in the region ($4\pi J+\alpha <2$) all the fugacities are relevant, and system is fully metallic. Other four regions are mixed phases where one or more fugacities become relevant thus implying special kinds of charge proliferation processes. These four phases therefore have partial or mixed order.

To find the fixed points of the second order equations, we first note the structure of the equation for $y_{+}$. Since, all the fugacities and $J$ are always positive, when $1+\frac{1}{z}-\alpha>0$, $y_{+}$ and $Jy_{1}y_{2}$ have to be zero. For Lorentz invariant fixed points corresponding to  $\frac{1}{z}=1$ and  the coupling constant space specified by $1+\frac{1}{z}-\alpha>0$, which we can also write as $A_{+}y_{+}^{2}/2+\alpha<2$, we find the following fixed point solution:
\begin{equation}
\text{FP1:} \; y_{1}=y_{2}=y_{+}=y_{-}=0, J=J^{*}, \alpha=\alpha^{*}<2 .
\end{equation}
When $1+\frac{1}{z}-\alpha<0$ i.e, $A_{+}y_{+}^{2}/2+\alpha>2$ , we get 
\begin{equation}
\text{FP2:}\;  y_{1}=y_{2}=y_{+}=y_{-}=0, J=J^{*}, \alpha=\alpha^{*}>2 .
\end{equation}
It is interesting to note that FP2 also corresponds to $\frac{1}{z}=1$. It is also worth emphasizing FP1 and FP2 describe different parts of the $J-\alpha$ plane. They are not critical points; there are lines of critical points in the surfaces containing these fixed points, which are the same as those found from the  first order RG equations.

Consider now $1+\frac{1}{z}-\alpha=0$. For the results to be sensible, $z$  must be positive and therefore  $\alpha>1$. The fixed points are now  non-Lorentz invariant, and they are given by
\begin{equation}
\text{FP3:}\;  y_{1}{*}=y_{2}^{*}=y_{-}^{*}=0, J^{*}=0,  A_{+}y_{+}^{*2}+2\alpha^{*}=4
\end{equation}
\begin{equation}
\text{FP4:}\;   y_{1}^{*}=y_{2}^{*}=y_{-}^{*}=0, A_{+}y_{+}^{*2}+2\alpha^{*}=4, J^{*}=\frac{1}{2}
\end{equation}
\begin{eqnarray}
\text{FP5:}\; &  &y_{2}^{*} = y_{-}^{*}=0,A_{+}y_{+}^{*2}+2\alpha^{*}=4 , J^{*}=\frac{\alpha^{*}}{\pi}, \nonumber \\ 
        &  & y_{1}^{*2}=\frac{(2-\alpha^{*})(\pi-2\alpha^{*})}{A_{1}\alpha^{*}}, 1<\alpha^{*}<\frac{\pi}{2}
\end{eqnarray}
These three sets of fixed points have continuously varying  dynamic scaling exponents. However, despite the presence of non-universal constants, $z$ is universally determined by $\alpha^{*}$. The only non-universality is in the location of the fixed points.  Importantly FP4 and FP5 are the sought after intermediate coupling fixed points controlling the cross-over between the local criticality and the global criticality, which are not accessible in weak coupling calculations. 

\section{The Phase Diagram}
In this section we construct the phase diagram, which is determined by FP1 and FP2. This is illustrated in Fig.2. In region, $(\pi J>2, \alpha >2)$ all the fugacities are irrelevant and hence  this phase is fully superconducting and is labeled as Superconducting. Here, each flavor of charges are bound in dipoles and the dipoles are formed between charges of opposite sign. Also, $y_{+}$ is irrelevant, implying  a quadrupolar order:  dipoles formed between two different flavors of charges, but of the same sign, are controlled by $y_{+}$ and hence in this region the inter-flavor dipoles are bound in quadrupoles. Following the similar argument for $y_{-}$, we can conclude that inter-flavor dipoles formed between charges of different flavors and opposite signs are bound as quadrupoles. These different dipolar and quadrupolar orders are shown in Fig.2. 
In the region, $(4\pi J+ \alpha <2)$, all the fugacities, $y_{1}, y_{2}, y_{+}, y_{-}$ are relevant,  and this is fully disordered or the metallic phase. Both flavors of charges proliferate and also the inter-flavor dipoles proliferate and we label this as Metallic.
In the region, $(\pi J>2, \alpha <2)$, only $y_{1},y_{2},y_{-}$ are irrelevant and only $y_{+}$ is relevant. This implies that each flavor of charges exhibit dipolar order but inter-flavor dipoles controlled by $y_{+}$ proliferate. Hence, this is a mixed phase. This phase is labeled as Mixed 1. So, the transition between the phases Mixed 1 and Superconducting is between a dipole gas and quadrupolar order.
In the region, $(\pi J<2, \alpha>2)$ ,$p_{2}$ charges are bound in dipoles and there is also quadrupolar order. But, $p_{1}$ charges proliferate. These facts are reflected by the irrelevance of $y_{2}, y_{+}, y_{-}$ and the relevance of $y_{1}$.  For this reason this is a phase of mixed order and we label this as Mixed 2.
In the region, $(\pi J+\alpha>2, \pi J<2, \alpha<2)$, $y_{2}, y_{-}$ are irrelevant, and $y_{1},y_{+}$ are relevant. $p_{2}$ charges are bound in dipoles and inter-flavor dipoles controlled by $y_{-}$ are bound as quadrupoles. But both the $p_{1}$ charges as well as the inter-flavor dipoles controlled by $y_{+}$ proliferate,  which is also an example of mixed order. This region is labeled as Mixed 3.
In the region, $4\pi J+\alpha>2, \pi J+\alpha<2$, only $y_{-}$ is irrelevant and all other fugacities are relevant. So, inter-flavor dipoles controlled by $y_{-}$ are bound as quadrupoles but $p_{1}, p_{2}$ charges and inter-flavor dipoles controlled by $y_{+}$ proliferate. This is also an example of mixed order and is labeled as Mixed 4.
\begin{figure}[htbp] 
   \centering
   \includegraphics[scale=0.5]{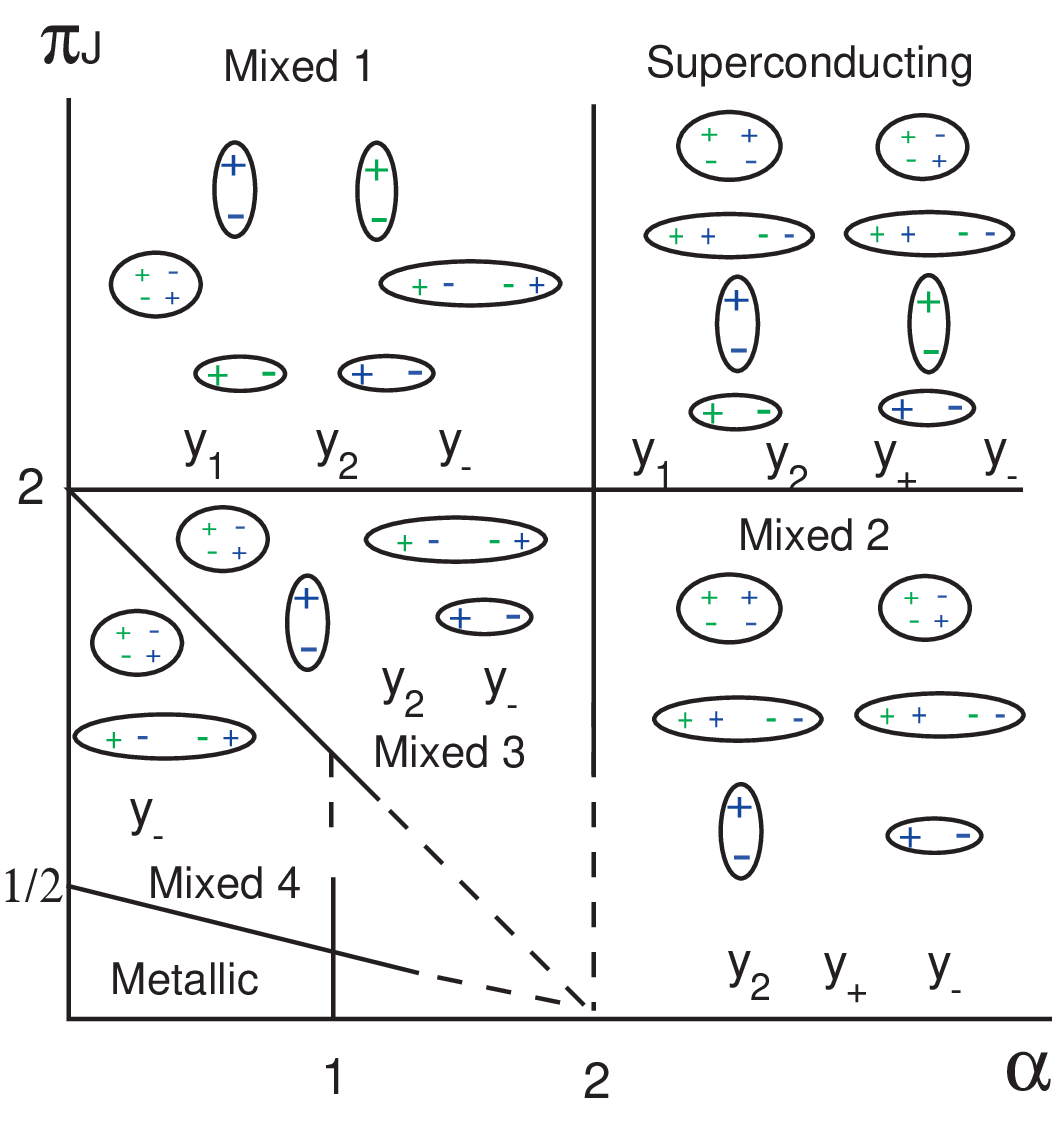} 
   \caption{Phase diagram. Other than  fully metallic and superconducting phases, there are four mixed phases characterized by the fugacities that are irrelevant. The strong coupling phase diagram should not be valid for $\pi J \ll 1$. The local critical boundary at $\alpha=1$, at weak coupling, requires a separate analysis, as described in the text.  }
   \label{fig:phase}
\end{figure}

\section{Conclusions}

In this section we note our observations in regard to the recent weak coupling analysis of the floating phase and  briefly describe the relevance of our theoretical work to various experiments that we have alluded to in the Introduction. In the near future, we hope to fully discuss the experimental consequences of our theory. For nanowires,  a number of important issues need to be addressed, such as  the effect of the external circuit for short wires, dissipative effects due to external circuit and the resistance of the phase slip cores,  boundary effects due to the leads \cite{Buchler, Subir2, Refael} and  inhomogenities \cite{Bezryadin4}. For experiments involving arrays, the real time dynamics and the current-voltage characteristics would be a major theoretical project. Presently, we can make only qualitative remarks that follow from our thermodynamic phase diagram.

\subsection{Floating phase and the local quantum criticality}
From the RG calculations at the second order, we found that $1/z$ depends on the fugacity $y_{+}$.
Therefore the locations of the nontrivial fixed points FP3, FP4 and FP5 found above 
depend on $y_{+}$. These fixed points can be accessed 
by tuning of $y_{+}$, in addition to tuning $J$ and $\alpha$. This is the reason
why they do not appear in the phase diagram, which is a cut in the  $J-\alpha$ plane. To recover the local quantum criticality observed in the weak coupling limit
we  note that  as $J$ is decreased below $2/\pi$, the fugacity $y_{1}$ becomes relevant, and we
can no longer use the flows derived from small fugacity expansions after a certain value of $J$. 
However, we observe that the growth of $y_{1}$ implies that $p_{1}$ charges  proliferate, and this fact allows us to
restrict $\phi_{1}\approx 0$. Because of the anisotroipc imaginary time interaction, 
$p_{2}$ charges may not proliferate at the same time. Using this approximation, the two-component sine-Gordon problem 
reduces now to a one-component sine-Gordon problem described by the action
\begin{equation}
S\approx \frac{1}{2}\sum_{{k,\omega}} \frac{2\pi}{\alpha}|\omega| |\phi_{2}|^{2}-2\frac{y_{2}}{a^{2}}\int d\vec{x} \cos(2\pi \phi_{2})
\end{equation}
Since, the propagator depends now only on $|\omega|$,  the problem is effectively zero dimensional.  Note that this action by itself has an approximate duality with the weak coupling action in terms of the junction variables
$\psi_{i}=\theta_{i}-\theta_{i+1}$ under the identification $y_{2}\to V$ and $\alpha\to 1/\alpha$. The self duality at  $\alpha=1$ is identical to the single junction problem. Thus, as 
the weak coupling limit is approached, at an intermediate coupling 
strength there is a change in the behavior from a global criticality to a local criticality.
This can be easily recognized  from the lack of self duality of the Josephson junction chain.
By a simple RG calculation we will get
\begin{eqnarray}
\frac{dy_{2}}{dl}&=&(1-\alpha)y_{2} \nonumber \\
\frac{d\alpha}{dl}&=&0
\end{eqnarray}
If $\alpha>1$, $y_{2}$ becomes irrelevant implying ordered state and as $\alpha<1$ the system becomes disordered.
Here the ordering is a local ordering of the individual junctions.

\subsection{Josephson junction array}
In this subsection we describe the relevance of our theoretical analysis regarding the experiments
on resistively shunted Josephson junction arrays in one dimension. We have found that the 
phase boundary between superconducting and fully metallic phase depends on the strength of the Josephson coupling of the superconducting grains. In our analysis we have predicted a complicated phase diagram involving mixed phases in addition to fully superconducting and fully metallic states,  depending on the strength of $J$. Only two transitions (between Mixed 1 and Superconducting and between Mixed 2 and Mixed 3) take place which are independent of the strength of $J$. Experimentally determined phase diagram due to Miyazaki {\em et al.}\cite{Miyazaki} confirms the strong coupling prediction that the phase boundary between ordered and disordered phase depends on $J$ and it also shows that in the weak coupling limit there is a part of the phase boundary at $\alpha=1$ which is independent of $J$ . More detailed experiments are necessary to establish the complete phase diagram on a quantitative level for the entire $J-\alpha$ plane and would be possible when we establish the signature of many of the mixed phases, which are mainly dynamical.   

Apart from the region in which the system is fully metallic, all other regions correspond to either fully superconducting and mixed phases.  We expect these superconducting and mixed phases to demonstrate power law behaviors for the temperature dependence of zero bias resistance: different mixed phases and the fully superconducting phase can be distinguished by their different temperature exponents. Similar power law behaviors are also expected for the I-V characteristic. These power law exponents will depend on $J$ and $\alpha$ which can be inferred from the linearized RG equations about the fixed points. 

\subsection{Superconducting nanowires} 
There is also a broad relevance of the dissipative Josephson junction array in the context of the experiments on superconducting nanowires.\cite{Bezryadin1,Bezryadin2,Bezryadin3,Bezryadin4,Chan} At $T=0$  quantum phase slips should play the key role in  determining if the system will be superconducting or resistive.  An experiment by Bezryadin {\em et al.}\cite{Bezryadin1} on  a superconconducting nanowire observed a dissipative phase transition similar to a single junction problem when the total normal resistance of the wire $R_{N}$ exceeded $R_{Q}$.  Later work\cite{Bezryadin2} implied that it is rather the resistance per unit length, hence the diameter of the wire, that is of significance.

Quantum phase slips are topological excitations in which the phase of the superconducting order parameter slips by a quantized amount and can be viewed as a vortex in the space-time manifold at $T=0$. If, for the moment, we can ignore dissipation and model the wire at $T=0$ as a $(1+1)$-dimensional XY-model, it is well known that such a system cannot exhibit a sharp phase transition (Kosterlitz-Thouless) without consideration of topology or vortex unbinding; smooth Gaussian fluctuations, ``spin waves'', cannot destroy superconductivity. The finite temperature properties in the proximity of the QCP can be understood in terms of universal scaling functions, and  for a finite wire, a finite size scaling analysis becomes necessary.

For a quantum system, the above picture must be modified because statics and dynamics are intricately intertwined. So, one must specify an appropriate dynamical model. In one such model Cooper pairs are allowed to disappear in a pool of normal electrons.\cite{Subir2} Thus, it is the actual phase of a superconducting grain that is coupled to an Ohmic heat-bath. While this may be sensible for a system of superconducting grains embedded in a normal metal.\cite{Larkin,Spivak}, it does not seem to reflect the physical situation in a superconducting nanowire. 

The RSJJA is another model for a superconducting nanowire, at least as far as the global $T=0$ phase diagram is concerned. The Josephson energy $V$ provides an energy barrier for phase slips for currents up to the critical current, and the topological excitations that are so essential are automatically built in the model. At a coarse grained level, on the scale of a coherence length $\xi$,  we can think of the wire to be partitioned into superconducting segments interrupted by the cores of the phase slips forming Josephson junctions. The characteristic frequency of the phase slips is given by the gap at which the conductivities of the normal metal and the superconductor are the same.   So, the system will have three coupling constants $\alpha$, $E_{0}$,  and $V$ corresponding to the degree of dissipation, the charging energy of the grains and the Josephson coupling strength. Here, the dissipation strength $\alpha=R_{Q}/R_{\xi}$ and $R_{\xi}=\rho \xi/\pi r^{2}$, where $\rho$ is the resistivity and  $r$ is the wire radius. Importantly, we believe that the correct model for the dissipative coupling is the one in which the phase {\em difference} across the phase slip is coupled to an Ohmic heat bath, as for RSJJA, unlike the model of Ref.~\onlinecite{Subir2}.

To determine the charging energy and the Josephson coupling strength, we consider the  continuum action that describes the Mooij-Sch\"{o}n\cite{Mooij} mode (gapless plasmon 
mode arising out of incomplete screening in one dimension) is given by
\begin{equation}
S=\frac{\mu}{2\pi}\int_{0}^{\beta} d\tau \int_{-L/2}^{L/2} dx[c_{s}(\partial_{x} \phi)^{2}+\frac{1}{c_{s}}(\partial_{\tau}\phi)^{2}]
\end{equation}
where $\mu$ and $c_{s}$ are the dimensionless kinetic admittance of 
the superconducting nanowire and the speed of propagation of the plasmon mode 
respectively. We have set $\hbar=1$. $\mu$ and $c_{s}$ are related to the 
superfluid density and the capacitance of the wire by the relations 
$ \frac{\mu c_{s}}{2\pi}=\frac{\rho_{s}}{2}=\frac{n_{s}A}{4m}$ and 
$\frac{\mu}{2\pi c_{s}}=\frac{\tilde{C}}{8e^{2}}$. $A$ is the of the cross-section, $\tilde{C}$
is the capacitance per unit length, and $n_{s}$ the bulk superfluid density. If we discretize this action on the scale of a  lattice spacing $a$,  
of the order of the superconducting coherence length $\xi$, we get the action 
of a Josephson junction chain with its parameters fixed by those of the 
Mooij-Sch\"{o}n mode. For the Josephson junction chain we get the following coupling constant 
relations: $V=\frac{\mu c_{s}}{\pi a}$ and $\frac{\mu a}{c_{s}}=\frac{1}{E_{0}}$. 
For the strong coupling limit of the Josephson junction array the dimensionless 
parameter $J=\mu/\pi$. If we define kinetic inductance per unit length of the wire 
as $\tilde{L}=1/e^{2}\rho_{s}$ we get $\mu=R_{Q}/\sqrt{\tilde{L}/\tilde{C}}$.
Kinetic inductance per unit length can be expressed in terms of London penetration depth
$\lambda_{L}$ as $\tilde{L}=\frac{\mu_{0}\lambda_{L}^{2}\xi}{Al_{0}}$ where $l_{0}$ is 
the mean free path of the electrons. Converting this in terms of the normal resistance per unit 
length of the wire $\tilde{R}$ we get $\tilde{L}\approx\frac{\hbar\tilde{R}}{1.76\pi k_{B}T_{c}}$
where we have restored $\hbar$.

We should note that, to leading order, we have $J\propto r$, as the capacitance per unit length has only a weak logarithmic dependence on $r$. So, a phase transition dependent on $J$ or $\mu$ implies a dependence on the radius of the wire,  that is, it tells us a critical radius beyond which superconductivity can be destroyed and these facts are in qualitative agreement with the results of Zaikin {\em et al.} \cite{Zaikin} and B\"{u}chler {\em et al.} \cite{Buchler}. This fact is also in qualitative agreement with the experimental results when the resistance per unit length is plotted against the temperature.\cite{Bezryadin2}  In recent experiments on single-crystal Sn nanowires , Tian {\em et al.} \cite{Chan} has also emphasized the role of the diameter of the wire for very low temperature measurements,  well below $T_{c}$.

Upto now it might seem that there are no differences between our results and those of Zaikin {\em et al.} and B\"{u}chler {\em et al}. If we go through the data of Lau {\em et al.} \cite{Bezryadin2} carefully, we will find that for the MoGe wires used in the experiments $\alpha>2$. So, in these experiments only control parameter is dimensionless admittance. This is the reason why our quantitative prediction $\mu=2$ is the same. But, significant differences will arise if the diameter and other system parameters can be adjusted such that $\alpha<2$; in this regime, we will observe transitions between the mixed phases and also between the metallic and the mixed phases, depending on the coupling strengths. In the mixed phases, the resistance will follow various power laws as a function of temperature. For, $\alpha<2$, these exponents will be depend on both $\mu$ and $\alpha$ in contrast to the situation when $\alpha>2$, where the exponent depends only on $\mu$. So, only if $\alpha$ as well as $\mu$ can be made sufficiently small, a local quantum criticality can be observed. Prediction of the new transitions is the outcome of treating the wire as an effective RSJJA. and we believe can be verified in more elaborate set of experiments on thinner wires and different materials.

Before concluding this section we should mention that we did not take into account a few effects. One is the boundary effects due to the leads \cite{Buchler, Subir2, Refael} and the second is the possibility of inhomogeneties resulting in weak links. That inhomogeneties can play an important  role is evident from the recent experiments of Bollinger {\em et al.} \cite{Bezryadin4} Another experimentally relevant issue is the effect of disorder on quantum phase slips. In a recent paper Khlebnikov and Pryadko \cite{Khlebnikov} considered the effect of disorder and demonstrated that disorder can bind the spatial coordinates of the phase slips and anti-phase slips and hence convert the problem to an effective $(0+1)$ dimensional problem. They found that the phase transition takes place at $\mu=1$ and belongs to the dissipative universality class which we have defined to be the local quantum criticality. We hope to return to these interesting effects relevant to the experiments in a future work.

\section{Acknowledgements}
We thank  M. Tinkham for discussions. A part of this work was carried out by one of us (S. C.) at the Aspen Center for Physics during the summer workshop ``Coherence and Dissipation in Quantum Systems'', 2004. This work was supported by a grant from the National Science Foundation: NSF-DMR 0411931.

 \appendix
 \section{\label{app:A}The Derivation of the Renormalization Group Equations}
In the small fugacity limit we can expand the partition function in  powers of fugacities and truncate at the quadratic order. If $\phi_{i}$ has frequency and momentum components in the  shell $a^{-1}e^{-l}<\omega<a^{-1}$ and $a^{-1}e^{-\frac{l}{z}}<k<a^{-1}$ , we define it to be the fast mode, $\phi_{i}^{f}$; otherwise, a mode is defined to be slow $\phi_{i}^{s}$. For a fixed frequency shell, we integrate out the the momenta, and then rescale both frequency and momentum in the resulting action for the slow part. 
\begin{figure}[htbp] 
   \centering
   \includegraphics[scale=0.5]{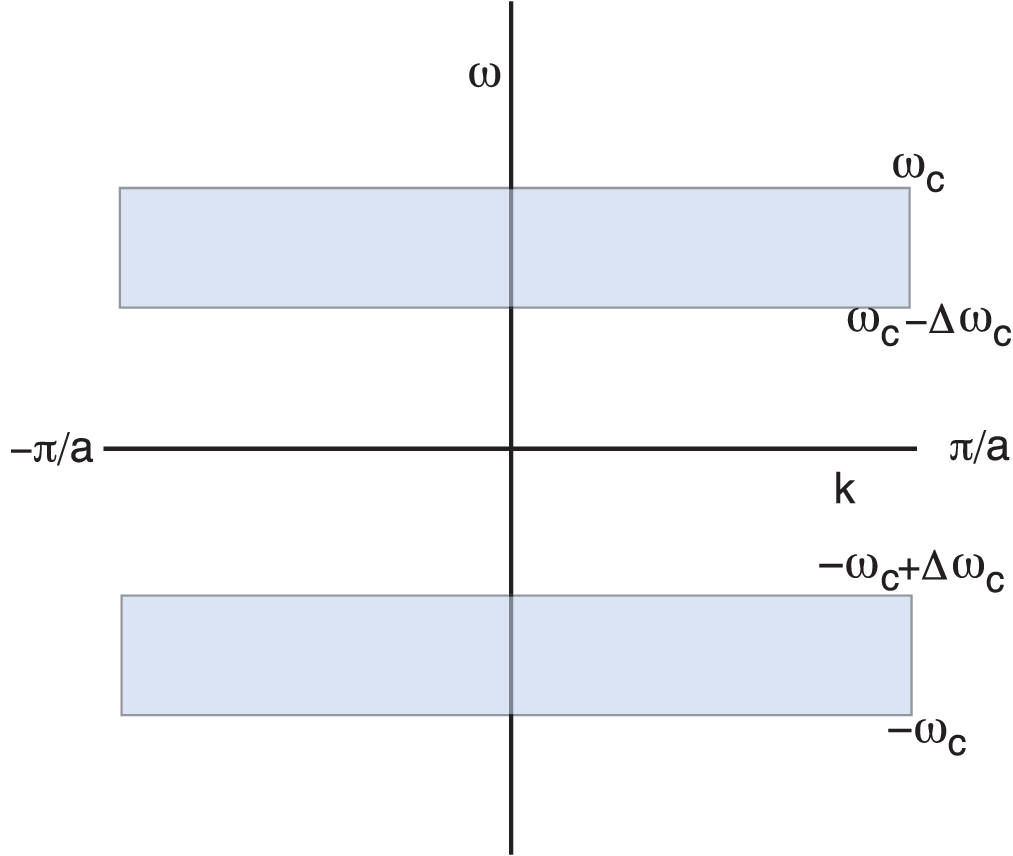} 
   \caption{Renormalization group shell integration. The integration proceeds by integrating out the shaded areas.}
   \label{fig:RG}
\end{figure}
The dynamic exponent $z$ is necessary to capture the effects of the anisotropic interaction.  The reason for this special order of integration  is because of $\frac{1}{|\omega|}$ in the propagator for $\phi_{2}$. If we reversed the order of integration,  we would have encountered spurious infrared divergence from the frequency integration. While rescaling we must maintain the periodicity of the cosine terms, which implies that the field renormalization constant is unity. Cosine averages are being calculated with respect to the fast degrees of freedom being integrated out
and it is denoted as $\langle .....\rangle^{f}$.  For any operator $O[\phi_{1},\phi_{2}]$, the average $\langle O\rangle^{f}$ is defined as
\begin{equation}
\langle O \rangle^{f}=\int D[\vec{\phi}^{f}]\exp[-S[\vec{\phi}^{f}]]O[\phi_{1},\phi_{2}]
\end{equation}
 We then arrive at the following action  action for the slow degrees of freedom:
\begin{widetext}
\begin{equation}
{\cal S}[\vec{\phi}_{s}]=\frac{1}{2}\sum_{k,\omega}\vec{\phi}^{\ast}_{s}\widetilde{G}^{-1}\vec{\phi}_{s}-\sum_{i}\frac{2y_{i}}{a^{2}}\int d\vec{x}\langle \cos(2\pi \phi_{i})\rangle^{f}  - \sum_{\eta=\pm}\frac{2y_{{\eta}}}{a^{2}}\int d\vec{x}\langle \cos(2\pi (\phi_{1} +\eta \phi_{2}))\rangle^{f} -\sum_{i,j=1}^{2}I_{ij}-\sum_{i=1,2,\alpha=\pm}I_{i \alpha}-\sum_{\alpha,\beta=\pm}I_{\alpha \beta}
\end{equation}
where $I_{ij}$, $I_{i\alpha}$ and $I_{\alpha \beta}$ represent the following second order contributions: 
\begin{eqnarray}
I_{i,j} & = &(\frac{2y_{i}y_{j}}{a^{4}})\int_{\vec{x}_{1},\vec{x}_{2}} \langle \cos(2\pi \phi_{i}(\vec{x}_{1}))\cos(2\pi \phi_{j}(\vec{x}_{2})) \rangle^{f} \nonumber \\
I_{i,\alpha}&=&(\frac{2y_{i}y_{\alpha}}{a^{4}})\int_{\vec{x}_{1},\vec{x}_{2}}\langle \cos(2\pi \phi_{1}(\vec{x}_{2})+2\pi \alpha \phi_{2}(\vec{x}_{2})) \cos(2\pi \phi_{i}(\vec{x}_{1}))\rangle^{f} \nonumber \\
 I_{\alpha,\beta} & = &(\frac{2y_{\alpha}y_{\beta}}{a^{4}})\int_{\vec{x}_{1},\vec{x}_{2}}\langle \cos(2\pi \phi_{1}(\vec{x}_{1})+2\pi \alpha \phi_{2}(\vec{x}_{1})) \cos(2\pi \phi_{1}(\vec{x}_{2})+2\pi \beta \phi_{2}(\vec{x}_{2}))\rangle^{f}
\end{eqnarray}
\end{widetext}

As previously used in the text, we are using the vector notation $\vec{x}=(x,\tau)$ and $\int_{\vec{x}}=\int dx d\tau$. The calculation involves the propagators $G_{ij}^{f}(x_{2}-x_{1},\tau_{2}-\tau_{1})\equiv G_{ij}^{f}(\vec{x}_{2}-\vec{x}_{1})=G_{ij}^{f}(\vec{r})$, which  has an oscillatory behavior for long distance due to the sharp cutoff in frequency . With suitable regularization we can avoid the complications that can arise due to this oscillatory behavior. Different regularizations lead to different nonuniversal constants in the recursion relations, but they do not affect the universal properties. We use the following regularized propagators
\begin{equation}
G_{11}^{f}(\vec{r}) =  \frac{J}{2\pi}K_{0}(\lambda \sqrt{|\vec{r}|^{2}+a^{2}}) =-G_{12}^{f}(\vec{r})=-G_{21}^{f}(\vec{r})
\end{equation}
and
\begin{equation}
G_{22}^{f}(\vec{r}) =  \delta(x)\frac{\alpha}{2\pi^{2}}K_{0}(\lambda \sqrt{\tau^{2}+a^{2}}) +\nonumber\\ \frac{J}{2\pi}K_{0}(\lambda \sqrt{|\vec{r}|^{2}+a^{2}})
\end{equation}
where $\lambda= a^{-1}e^{-l}$. $K_{0}$ is the modified Bessel function. For distance much greater than the cutoff these regularized propagators fall off exponentially and for distance much less than $\lambda^{-1}$ they have logarithmic dependence on distance. For example
\begin{widetext}
\begin{eqnarray}
\frac{y_{i}}{a^{2}} \int d \vec{x} \langle \cos(2 \pi \phi_{i}(\vec{x})) \rangle^{f} 
=\frac{y_{i}}{a^{2}} \int d \vec{x} \langle \cos(2 \pi \phi_{i}^{s}(\vec{x})+2 \pi \phi_{i}^{f}(\vec{x})) \rangle^{f} 
& &= \frac{y_{i}}{a^{2}}\int d \vec{x} \cos(2 \pi \phi_{i}^{s}(\vec{x})) \exp[-2 \pi^{2}\langle \phi_{i}^{2} (\vec{x})\rangle^{f}] \nonumber \\
& & = \frac{y_{i}}{a^{2}}\exp[-2\pi^{2}G_{ii}^{f}(0)]\int d \vec{x} \cos(2 \pi \phi_{i}^{s}(\vec{x})) 
\end{eqnarray}
\end{widetext}
After rescaling rescaling the coordinates we obtain
\begin{equation}
y_{i}(l)=\exp[(1+\frac{1}{z})l-2\pi^{2}G_{ii}^{f}(0)]y_{i}(0)
\end{equation}
When we compute an average $\langle \cos(2\pi \phi_{1}(\vec{x})\pm 2\pi \phi_{2}(\vec{x}))\rangle^{f}$ it involves computing the average $\langle (\phi_{1}(\vec{x})\pm \phi_{2}(\vec{x}))^{2}\rangle^{f}$. Due to the mutual interaction between $\phi_{1}$ and $\phi_{2}$ fields we obtain
\begin{eqnarray}
& &\langle (\phi_{1}(\vec{x})\pm\phi_{2}(\vec{x}))^{2}\rangle^{f} \nonumber \\
& &=G_{11}^{f}(0)+G_{22}^{f}(0)\pm2G_{12}^{f}(0)
\end{eqnarray}
which leads to 
\begin{equation}
y_{\pm}(l)=\exp[(1+\frac{1}{z})l-2\pi^{2}(G_{11}^{f}(0)+G_{22}^{f}(0)\pm 2G_{12}^{f}(0))]y_{\pm}(0)
\end{equation}
From equation (A5) and (A7) we obtain the first order RG equations for the fugacities. 

Now we illustrate the calculations of the second order terms. All the second order terms involve two space-time coordinates $(\vec{x}_{1},\vec{x}_{2})$ but the propagators in real space-time are only functions of relative space-time coordinate $(\vec{x}_{2}-\vec{x}_{1})$. For this reason we do a coordinate transformation to the center of mass coordinate $(\vec{R}=\frac{\vec{x}_{1}+\vec{x}_{2}}{2})$ and the relative coordinate $(\vec{r}=\vec{x}_{2}-\vec{x}_{1})$. The second order contribution $I_{11}$  will involve only the isotropic propagator $G_{11}^{f}(|\vec{r}|)$. In order to calculate such a term we can go over to polar coordinates for the relative vector $\vec{r}$ which has following three ranges
\begin{equation}
\int d\vec{r} = \int_{0}^{2\pi} d\theta[\int_{0}^{a} rdr +\int_{a}^{ae^{l}} rdr +\int_{ae^{l}}^{\infty} rdr]
\end{equation}
In the first range of integration we can take the propagator $G_{11}^{f}(|\vec{r}|)=G_{11}^{f}(0)$; in the second range either we can set $G_{11}^{f}(|\vec{r}|)=G_{11}^{f}(a)$ or carry out the integral with the regularized logarithmic form of the propagator combined with a gradient expansion of the cosine terms. With suitable regularization, $G_{11}^{f}(|\vec{r}|)$ can be converted into a short range function of $|\vec{r}|$ and third part of the integral can be made negligible.\cite{Kogut, Ohta} The specific regularization controls how fast and how smooth the propagator falls off with the distance but does not affect the universal properties. This is what we achieve with the regularized propagator mentioned above.

Due to the anisotropic part in the propagator $G^{f}_{22}$,  the above decomposition into polar coordinates can be incorrect for the second order terms involving the field $\phi_{2}$. So, we have to check  all the second order terms mentioned above independently. As we will demonstrate below the second order terms consisting of cross correlation of the fields $\phi_{1}, \phi_{2}$ lead to integrals over combination of $G_{11}^{f}, G_{22}^{f}$ and $G_{12}^{f}$. If this combination turns out to be isotropic in space-time we can proceed with polar coordinate decomposition. Otherwise we need to follow a tricky decomposition which we will use below for $I_{22}$. But, there is an interesting aspect of these second order contributions. If the second order term involves product of two different fugacities, it leads to relevant contributions only if the combinations of the fields are contracted at the same space-time point which renormalizes other fugacities. This aspect simplifies the calculation of the second terms involving product of two different fugacities. 

For illustrative purposes of the above procedures first we will pick $I_{11}$.
\begin{widetext}
\begin{eqnarray}
I_{11} &=&\frac{y_{1}^{2}}{2a^{4}}\int_{\vec{R},\vec{r}}\langle e^{i2 \pi(\phi_{1}(\vec{R}+\vec{r}/2)+  \phi_{1}(\vec{R}-\vec{r}/2))}+c.c. 
           +e^{i2 \pi(\phi_{1}(\vec{R}+\vec{r}/2)- \phi_{1}(\vec{R}-\vec{r}/2))} +c.c.\rangle^f \nonumber  \\
            &=&\frac{y_{1}^{2}}{a^{4}}e^{-4 \pi^{2}G_{11}^{f}(0)}\int_{\vec{R},\vec{r}}[(e^{-4\pi^{2}G_{11}^{f}(\vec{r})}-1) 
            \cos(2\pi (\phi_{1}(\vec{R}+\vec{r}/2)+ \phi_{1}(\vec{R}-\vec{r}/2))) \nonumber \\
            &  &+(e^{4\pi^{2}G_{11}^{f}(\vec{r})}-1)\cos(2\pi( \phi_{1}(\vec{R}+\vec{r}/2)
           - \phi_{1}(\vec{R}-\vec{r}/2))) ]                
 \end{eqnarray}  
 \end{widetext}         
First integral generates higher harmonic $\cos(4\pi \phi_{1}(\vec{R}))$ from the first integral range of  the 
relative coordinate. The same integral in the second integral range generates a term proportional to the higher harmonic multiplied by square of the gradient of the field $\phi_{1}$. So, obviously all these terms are irrelevant. Relevant contribution comes from the second integral when we concentrate on the second integral range of the relative coordinate.  After a gradient expansion for cosine term we obtain,
\begin{equation}
I_{11} \approx  -\frac{A_{1}}{2}y_{1}^{2}\int d\vec{x}(\nabla \phi_{1})^{2}
\end{equation}
$A_{1}$ is a regularization dependent constant which in our case is $4\pi^{3}$. This will contribute to the renormalization of $J$. Now we will pick a cross term e.g.
\begin{widetext}
\begin{eqnarray}
I_{12} &=& \frac{y_{1}y_{2}}{2a^{4}}\int_{\vec{R},\vec{r}}\langle e^{i(2 \pi\phi_{1}(\vec{R}+\vec{r}/2)+ 2 \pi\phi_{2}(\vec{R}-\vec{r}/2))}+c.c. 
            +e^{i(2 \pi\phi_{1}(\vec{R}+\vec{r}/2)-2 \pi\phi_{2}(\vec{R}-\vec{r}/2))} +c.c.\rangle^f \nonumber  \\
            &=&\frac{y_{1}y_{2}}{a^{4}}e^{-2 \pi^{2}(G_{11}^{f}(0)+G_{22}^{f}(0))}\int_{\vec{R},\vec{r}}[(e^{-4\pi^{2}G_{12}^{f}(\vec{r})}-1)
            \cos(2\pi \phi_{1}(\vec{R}+\vec{r}/2)+2\pi \phi_{2}(\vec{R}-\vec{r}/2))  \nonumber \\
            &  &+(e^{4\pi^{2}G_{12}^{f}(\vec{r})}-1)\cos(2\pi \phi_{1}(\vec{R}+\vec{r}/2)
            -2\pi \phi_{2}(\vec{R}-\vec{r}/2)) ]                
 \end{eqnarray} 
 \end{widetext}          
As mentioned above here we have to deal with only an isotropic propagator $G_{12}^{f}$ and hence polar coordinate will be useful. From the first range of the $\vec{r}$ integral we will get the contribution 
\begin{eqnarray}
I_{12} &\approx& (\frac{y_{1}y_{2}}{a^{2}})e^{-2 \pi^{2}(G_{11}^{f}(0)+G_{22}^{f}(0))}(\pm 4\pi^{2}G_{12}^{f}(0)) \times \nonumber \\
            &              &\int_{\vec{R}}\cos(2\pi \phi_{1}(\vec{R})\mp2\pi \phi_{1}(\vec{R}))
\end{eqnarray}
 which renormalizes $y_{\pm}$. In the second range of the integrals we will do a gradient expansion for the cosine terms which leads to the contributions proportional to $cos(2\pi \phi_{1}(\vec{R})\pm 2\pi \phi_{2}(\vec{R}))\times (\nabla_{R}\phi_{1}\mp \nabla_{R}\phi_{2})^{2}$. After Fourier transformation this leads to the terms which are combinations of frequency or momentum with cosine and hence irrelevant.
 
Two interesting second order terms are $I_{1+}$ and $I_{1-}$. For relevant part of these terms we get
\begin{widetext}
\begin{eqnarray}
I_{1+} &\approx& \frac{y_{1}(l)y_{+}(l)}{a^{4}}e^{-2(1+1/z)l}\int_{\vec{R},\vec{r}}(e^{4\pi^{2}(G_{11}^{f}(\vec{r})+G_{12}^{f}(\vec{r}))}-1)\cos(2\pi \phi_{1}(\vec{R}-\vec{r}/2)+2\pi \phi_{2}(\vec{R}-\vec{r}/2)-2\pi \phi_{1}(\vec{R}+\vec{r}/2))\nonumber \\
I_{1-} &\approx&\frac{y_{1}(l)y_{-}(l)}{a^{4}}e^{-2(1+1/z)l}\int_{\vec{R},\vec{r}}(e^{4\pi^{2}(G_{11}^{f}(\vec{r})-G_{12}^{f}(\vec{r}))}-1)\cos(2\pi \phi_{1}(\vec{R}+\vec{r}/2)+2\pi \phi_{2}(\vec{R}-\vec{r}/2)-2\pi \phi_{1}(\vec{R}-\vec{r}/2))\nonumber
\\
\end{eqnarray}
\end{widetext}
Recalling that $G_{11}^{f}+G_{12}^{f}=0$ we can see that relevant part of $I_{1+}$ vanishes identically.
As both $G_{11}^{f}$ and $G_{12}^{f}$ are isotropic in space time polar decomposition is valid for $I_{1-}$. Finally we get 
\begin{equation}
I_{1-} \approx  \frac{y_{1}y_{-}}{a^{2}} 4\pi Jl\int d\vec{x} \cos(2 \pi \phi_{2})
\end{equation}
$I_{11}, I_{12}, I_{1+}$ and $ I_{1,-}$ exhaust all the second order terms that involve the fugacity $y_{1}$. Now we will also exhaust all the second order terms that involve the fugacity $y_{2}$.
For the relevant part of $I_{2+}$ and $I_{2-}$ we get 
\begin{widetext}
\begin{eqnarray}
I_{2+} &\approx& \frac{y_{2}(l)y_{+}(l)}{a^{4}}e^{-2(1+1/z)l}\int_{\vec{R},\vec{r}}(e^{4\pi^{2}(G_{22}^{f}(\vec{r})+G_{12}^{f}(\vec{r}))}-1)\cos(2\pi \phi_{1}(\vec{R}-\vec{r}/2)+2\pi \phi_{2}(\vec{R}-\vec{r}/2)-2\pi \phi_{2}(\vec{R}+\vec{r}/2))\nonumber \\
I_{2-} &\approx&\frac{y_{2}(l)y_{-}(l)}{a^{4}}e^{-2(1+1/z)l}\int_{\vec{R},\vec{r}}(e^{4\pi^{2}(G_{22}^{f}(\vec{r})-G_{12}^{f}(\vec{r}))}-1)\cos(2\pi \phi_{2}(\vec{R}+\vec{r}/2)+2\pi \phi_{1}(\vec{R}-\vec{r}/2)-2\pi \phi_{2}(\vec{R}-\vec{r}/2))\nonumber
\\
\end{eqnarray}
\end{widetext}
Both of the combinations $G_{22}^{f}(\vec{r})\pm G_{12}^{f}(\vec{r})$ involve anisotropic part of the propagator $G_{22}^{f}$. Since, the relevant contributions from the two terms above will come from the contraction of the fields at the same space-time point we can set $\vec{r}=0$ without further difficulty
and obtain
\begin{eqnarray}
I_{2+}& \approx & \frac{y_{2}y_{+}}{a^{2}} 2\alpha l\int d\vec{x} \cos(2 \pi \phi_{1}) \nonumber \\
I_{2-}& \approx & \frac{y_{2}y_{-}}{a^{2}} 2(\alpha +2\pi J)l\int d\vec{x} \cos(2 \pi \phi_{1})
\end{eqnarray}
Finally we consider $I_{22}$ which will exhaust all the terms that involve $y_{2}$.
\begin{widetext}
\begin{eqnarray}
I_{22}&=&\frac{y_{2}^{2}}{a^{4}}e^{-4 \pi^{2}G_{22}^{f}(0)}\int_{\vec{R},\vec{r}}[(e^{-4\pi^{2}G_{22}^{f}(\vec{r})}-1) 
            \cos(2\pi \phi_{2}(\vec{R}+\vec{r}/2)+2\pi \phi_{2}(\vec{R}-\vec{r}/2))  \nonumber \\
            &  &+(e^{4\pi^{2}G_{22}^{f}(\vec{r})}-1)\cos(2\pi \phi_{2}(\vec{R}+\vec{r}/2)
           -2\pi \phi_{2}(\vec{R}-\vec{r}/2)) ]                
 \end{eqnarray}  
 \end{widetext}     
Regardless of the anisotropic interaction the first integral leads to higher harmonics of $\phi_{2}$ and hence can be ignored. Calculation of the second integral is tricky but leads to contributions involving $(\partial_{\tau} \phi_{2})^{2}$ and $(\partial_{x} \phi_{2})^{2}$. But, in the original action such terms are not present and these are also of higher order than $|\omega | | \phi_{2}|^{2}$ and hence irrelevant. Though the second integral produces irrelevant terms we will describe the way to calculate it as such a calculation will be required later on for $I_{--}$. We will break the integral into two parts corresponding to only onsite contraction and offsite contraction of the field $\phi_{2}$. 
\begin{widetext}
\begin{eqnarray}  
I_{22} & \approx &\frac{y_{2}^{2}}{a^{4}}e^{-4 \pi^{2}G_{22}^{f}(0)}\int_{\vec{R},\vec{r}}(e^{4\pi^{2}G_{22}^{f}(\vec{r})}-1)\cos(2\pi \phi_{2}(\vec{R}+\vec{r}/2)  -2\pi \phi_{2}(\vec{R}-\vec{r}/2))\nonumber \\
           & = &\frac{y_{2}^{2}}{a^{3}}e^{-4 \pi^{2}G_{22}^{f}(0)}\int dx du d\tau (e^{4\pi^{2}G_{22}^{f}(0,\tau)}-1)\cos(2\pi \phi_{2}(x,u+\tau/2)  -2\pi \phi_{2}(x,u-\tau/2))\nonumber \\
           &    &+\frac{y_{2}^{2}}{a^{4}}e^{-4 \pi^{2}G_{22}^{f}(0)}\int_{\vec{R},\vec{r}}(e^{4\pi^{2}G_{22}^{f I}(|\vec{r}|)}-1)\cos(2\pi \phi_{2}(\vec{R}+\vec{r}/2)-2\pi \phi_{2}(\vec{R}-\vec{r}/2))\nonumber \\
           &    &-\frac{y_{2}^{2}}{a^{3}}e^{-4 \pi^{2}G_{22}^{f}(0)}\int dx du d\tau (e^{4\pi^{2}G_{22}^{f I}(0,\tau)}-1)\cos(2\pi \phi_{2}(x,u+\tau/2)  -2\pi \phi_{2}(x,u-\tau/2))
\end{eqnarray}
\end{widetext}
where $u=\frac{\tau_{1}+\tau_{2}}{2}$ and $\tau=\tau_{2}-\tau_{1}$. We have also broken $G_{22}^{f}$ into two parts $G_{22}^{fI}$ and $G_{22}^{fA}$ corresponding to isotropic and anisotropic interactions respectively. Rearrangement of $I_{22}$ in this form has been used previously by Bobbert et al.\cite{Bobbert1} For the first and third integrals which involve only onsite contraction we can break up the integrals into three parts
\begin{equation}
\int d\tau= \int_{o}^{a}d\tau +\int_{a}^{ae^{l}}d\tau +\int_{ae^{l}}^{\infty}d\tau
\end{equation}
Second range of the integrals contribute terms proportional to $(\partial_{u} \phi_{2})^{2}$. It turns out that the contributions from the first and third integrals cancel each other. Second integral involves only isotropic function of $|\vec{r}|$ and we can use polar coordinates as mentioned above which will contribute a term proportional to $(\nabla_{R}\phi_{2})^{2}$. 

Finally we have to consider only two other second order terms $I_{++}$ and $I_{--}$. Relevant part of these are given by
\begin{widetext}
\begin{eqnarray}
I_{++} &\approx & \frac{y_{+}^{2}(l)}{a^{4}}e^{-2(1+1/z)l}\nonumber \\
&\times&\int_{\vec{R},\vec{r}}(e^{4\pi^{2}(G_{11}^{f}(\vec{r})+G_{22}^{f}(\vec{r})+2G_{12}^{f}(\vec{r}))}-1)\cos(2\pi (\phi_{1}(\vec{R}+\vec{r}/2)+ \phi_{2}(\vec{R}+\vec{r}/2)- \phi_{1}(\vec{R}-\vec{r}/2)- \phi_{2}(\vec{R}-\vec{r}/2)))\nonumber \\
I_{--} &\approx & \frac{y_{-}^{2}(l)}{a^{4}}e^{-2(1+1/z)l} \nonumber \\
&\times&\int_{\vec{R},\vec{r}}(e^{4\pi^{2}(G_{11}^{f}(\vec{r})+G_{22}^{f}(\vec{r})-2G_{12}^{f}(\vec{r}))}-1)\cos(2\pi (\phi_{1}(\vec{R}+\vec{r}/2)- \phi_{2}(\vec{R}+\vec{r}/2)- \phi_{1}(\vec{R}-\vec{r}/2)+ \phi_{2}(\vec{R}-\vec{r}/2)))\nonumber \\
\end{eqnarray}
\end{widetext}
$G_{11}^{f}(\vec{r})+G_{22}^{f}(\vec{r})+2G_{12}^{f}(\vec{r})= \delta(x)\frac{\alpha}{\pi^{2}}K_{0}(\lambda \sqrt{\tau^{2}+a^{2}})$ which depends only on time. This implies for $I_{++}$ the relevant contribution is obtained by contracting the combination of the fields $\phi_{1}+ \phi_{2}$ at same spatial point but at different times. So, we need to integrate over only $\tau=\tau_{2}-\tau_{1}$. After breaking up the integral into three parts as we have done before the relevant contribution will come from the second range of the integral and we obtain
\begin{equation}
I_{++} \approx  -\frac{A_{+}}{2}y_{+}^{2}\int d\vec{x}(\partial_{\tau} \phi_{1})^{2}
\end{equation}
$A_{+}$ is a regularization dependent constant and for the regularization used above, $A_{+}=4\pi^{2}$.
We have ignored the irrelevant terms like $(\partial_{\tau} \phi_{2})^{2}$ and $\phi_{1}\phi_{2}$ involving spatial or temporal derivatives.

Since, $G_{11}^{f}(\vec{r})+G_{22}^{f}(\vec{r})+2G_{12}^{f}(\vec{r})$ involves both isotropic and anisotropic parts, the evaluation of $I_{--}$ is more complicated. But, we can carry out the calculation using the trick mentioned above for $I_{22}$. Ignoring irrelevant contributions we get
\begin{equation}
I_{--}\approx -\frac{A_{-}}{2}y_{-}^{2}\int d\vec{x}(\nabla \phi_{1})^{2}
\end{equation}
where $A_{-}$ is another regularization dependent constant and for the regularization method  chosen above $A_{-}=4\pi^{3}$. 

Since, we are integrating out the momenta for a fixed frequency shell, the coefficient of $\int d\vec{x} (\partial_{\tau}\phi_{1})^{2}$ has to be kept fixed. Due to this reason after collecting the second order terms involving $\int d\vec{x} (\partial_{\tau}\phi_{1})^{2}$ and rescaling the space-time coordinates we obtain,
\begin{equation}
\frac{1}{J^{'}}=e^{(\frac{1}{z}-1)l}[\frac{1}{J}+(A_{1}y_{1}^{2}+A_{+}y_{+}^{2}+A_{-}y_{-}^{2})l]
\end{equation}
which leads to
\begin{equation}
\frac{dJ}{dl}=J[(1-\frac{1}{z})-J(A_{1}y_{1}^{2}+A_{+}y_{+}^{2}+A_{-}y_{-}^{2})]
\end{equation}
The anisotropic scaling prescription, $k\to ke^{-l/z}$ can be used to keep the coefficient of $k^{2}$ fixed and $1/z$ is determined by the difference of the coefficients of the terms involving $\int d\vec{x} (\partial_{\tau}\phi_{1})^{2}$ and $\int d\vec{x} (\partial_{x}\phi_{1})^{2}$ generated at the second order RG transformation. This trick to extract the dynamic exponent in a perturbative analysis is well known.\cite{Senthil} Following this trick we get 
\begin{equation}
2(1-\frac{1}{z})=A_{+}y_{+}^{2}
\end{equation}
This completes the derivation of the recursion relations of the coupling constants upto the second order.

\end{document}